\documentstyle[12pt]{article}
\begin{document}
\def\mettresous#1\sous#2{\mathrel{\mathop{\kern0pt #2}\limits_{#1}}}
\def\sqr#1#2{{\vcenter{\vbox{\hrule height.#2pt
          \hbox{\vrule width.#2pt height#1pt \kern#1pt
           \vrule width.#2pt}
           \hrule height.#2pt}}}}
\def\square{\mathchoice\sqr68\sqr68\sqr{4.2}6\sqr{3}6}
\def\lrpartial{\mathrel{\partial\kern-.75em\raise1.75ex\hbox{
$\leftrightarrow$}}}
\def\om{\omega	}
\def\la{\lambda}

\def\lr #1{\mathrel{#1\kern-.75em\raise1.75ex\hbox{$\leftrightarrow$}}}

\def\p {\prime}
\def\dd {\dagger}
\def\ga {\alpha}
\def\gb {\beta}
\def\gg {\gamma}
\def\gd {\delta}
\def\ge {\epsilon}
\def\gta {\eta}
\def\gve {\var epsilon}
\def\gph {\phi}
\def\gvf {\varphi}
\def\gch {\chi}
\def\gth {\theta}
\def\gvth {\vartheta}
\def\gi {\iota}
\def\gk {\kappa}
\def\gl {\lambda}
\def\gm {\mu}
\def\gn {\nu}
\def\go {\omega}
\def\gp {\pi}
\def\gvp {\varpi}
\def\gr {\rho}
\def\gps {\psi}
\def\gvr {\varrho}
\def\gs {\sigma}
\def\gvs {\varsigma}
\def\gt {\tau}
\def\gx {\xi}
\def\gy {\upsilon}
\def\gz {\zeta}

\def\gG {\Gamma}
\def\gD {\Delta}
\def\gPh {\Phi}
\def\gTh {\Theta}
\def\gL {\Lambda}
\def\gO {\Omega}
\def\gP {\Pi}
\def\gPs {\Psi}
\def\gR {\Rho}
\def\gS {\Sigma}
\def\gY {\Upsilon}

\def\mettresous#1\sous#2{\mathrel{\mathop{\kern0pt #2}\limits_{#1}}}
\def\sqr#1#2{{\vcenter{\vbox{\hrule height.#2pt
          \hbox{\vrule width.#2pt height#1pt \kern#1pt
           \vrule width.#2pt}
           \hrule height.#2pt}}}}
\def\square{\mathchoice\sqr68\sqr68\sqr{4.2}6\sqr{3}6}
\def\ket#1{|#1\rangle}
\def\bra#1{\langle #1|}
\def\braket#1#2{\mathrel{\langle #1|#2\rangle}}
\def\elematrice#1#2#3{\langle #1|#2|#3 \rangle}
\def\inoutexpect#1{\elematrice{0,\mbox{out}}{#1}{0,\mbox{in}}}
\def\inout{\langle 0,\mbox{out}|0,\mbox{in}\rangle}
\def\lrpartial{\mathrel{\partial\kern-.75em\raise1.75ex\hbox{$\leftrightarrow$}}}
\def\lrD{\mathrel{{\rm D}\kern-.75em\raise1.75ex\hbox{$\leftrightarrow$}}}

\begin{flushright}
ULB-TH 94/19\\ November 1994\\
\end{flushright} \vskip 1.5 truecm
\centerline{\bf{The Semi-Classical Back Reaction to Black Hole Evaporation}}
 \vskip 1. truecm \centerline{  S. Massar\footnote{e-mail:
smassar @ ulb.ac.be}$\/^{,}$ \footnote{Boursier IISN}} \centerline{Service de
Physique Th\'eorique, Universit\'e Libre de Bruxelles,} \centerline{Campus
Plaine, C.P. 225, Bd du Triomphe, B-1050 Brussels, Belgium}
\vskip 1.5 truecm
{\bf Abstract }
The semi-classical back reaction to
black hole evaporation (wherein the renormalized energy momentum tensor
is taken as source of Einstein's equations) is analyzed in detail. It is
proven that the mass of a Schwarzshild black hole decreases according to
Hawking's law $dM/dt = - C/ M^2$ where $C$ is a constant of order one and
that the particles are emitted with a thermal spectrum at temperature $1/8\pi
M(t)$.

\vfill \newpage

On the basis of  Hawking's \cite{H}  derivation of black hole
radiation, wherein matter fields are quantized on the
fixed background of a collapsing star, the mass of a
black hole is expected to decrease according to the law
\begin{equation}
{dM\over dt}= -{C \over M^2}
\label{one}
\end{equation}
where $C$ is a constant of order one which takes into account the number
and spin of massless fields, the particles emitted at time $t$ being
distributed in a thermal spectrum with temperature \begin{equation}
T_H(t) = {1\over 8 \pi M(t)}
\label{two}
\end{equation}

Since Hawking's discovery much
interest has focused on the gravitational back reaction in order to confirm
or infirm (\ref{one}) and (\ref{two}) and hopefully in the process
learn something about quantum gravity.
One of the simplest
schemes in which to investigate the back reaction is the
semi-classical theory wherein the renormalized energy momentum
tensor is taken as source of
Einstein's equations $G_{\mu\nu} = 8 \pi \langle T_{\mu\nu} \rangle_{ren}$,
ie. the metric remains classical but follows self consistently the mean
energy of the field.

In this article we show that (\ref{one}) and (\ref{two}) are indeed
verified in the semi classical theory of black hole evaporation. However we
mention at the outset that due to the exponential Doppler shift experienced by
the field in its voyage from inside the star to ${\cal I}^+$ the semi
classical back reaction is not a consistent  approximation to the true back
reaction. This is not to say that the predictions of the semi classical
theory are necessarily wrong, but the detailed mechanism by which vacuum
fluctuations are converted into quanta will necessarily be totaly different
in a fully quantum theory in which the metric as well as the matter fields
fluctuate.

This article is inspired by
the work of Hajicek and Israel \cite{HI} and Bardeen \cite{B} who showed that
the semi classical back reaction is not incompatible with equation
(\ref{one}) and by the work of Parentani and Piran \cite{PP} who showed
numerically that (\ref{one}) is verified in a model in which the
renormalized energy momentum tensor takes a particularly simple form.

 We shall work in Bardeen's coordinates:
\begin{equation}
ds^2=-e^{2 \psi}(1-2m/r)dv^2 + 2 e^\psi dvdr
+r^2 d\Omega^2
\label{c1}
\end{equation}
In these coordinates Einstein's
equations are
\begin{eqnarray}
{\partial m \over \partial v} &=& 4 \pi r^2 T^r_{\
v}\nonumber\\
{\partial m \over \partial r} &=& -4 \pi r^2 T^v_{\
v}\nonumber\\
{\partial \psi \over \partial r} &=& 4 \pi r
T_{rr}\label{c2}
\end{eqnarray}
Note that $\psi$ is defined only up to the addition of an arbitrary
function of $v$ corresponding to a reparametrization of the $v$ coordinate.

We shall suppose that the right hand side of (\ref{c2}) is given by the
renormalized energy momentum tensor (for a review see
\cite{BD}). The advantage of the coordinate system (\ref{c1}) is that the
Hawking flux is encoded near the horizon in a negative energy flux
$T^r_{v}<0$. Were one to neglect the other components, the solution near the
horizon would be a simple Vaidya type metric with only parameter $m(v)= \int^v
\! dv \ T_{vv}$.

We shall proceed in three steps: first we shall suppose
that the renormalized energy momentum tensor resembles the renormalized
energy momentum tensor in the absence of back reaction. We shall then show
that under this hypothesis the metric coefficients in
(\ref{c1}) are slowly varying functions of $r$ and $v$. Finally we shall solve
adiabatically the Klein Gordon equation in this slowly varying metric and
show that the renormalized energy momentum tensor indeed posseses the
properties supposed at the outset thereby proving that the calculation is
consistent.

Our first task is to obtain estimates for $T_{\mu\nu}$ both far and near
the black hole. We begin with the former. We suppose that when $r$
is equal to a few times $2m$ (say $r=O(6m)$) there is only an outgoing flux
$T_{uu}(r>>2m)=L_H(u)/4\pi r^2$ where $L_H $ is the luminosity of the black
hole. This is
justified since in the absence of back reaction the other components of
$T_{\mu\nu}$ decrease as large inverse powers of $r$, for instance the trace
anomaly decreases as $m^2/ r^6$. We shall also suppose that $L_H$ is small and
varies slowly. Hence when $r>O(6m)$ an outgoing Vaidya metric is an exact
solution of Einstein's equations \begin{eqnarray} &ds^2 = -(1-{2 m(u) \over
r})du^2 -2 du dr +r^2 d\omega^2&\nonumber\\ &m(u) = \int^u \! du \
L_H(u)&\label{c3} \end{eqnarray}
The change of coordinates from (\ref{c2}) to (\ref{c3})
is obtained by writing the equation for infalling radial null geodesics in
the metric (\ref{c3}) as
\begin{equation}
F dv = du +  {2 dr\over 1 - 2 m(u)/r}\label{cIV}
\end{equation}
where $F$ is an integration factor. Upon using (\ref{cIV}) to change
coordinates from the set $(u,r)$ to $(v,r)$ one
finds that $e^{\psi}=F$ and $m(r,v)=m(u)$.
Hence when $r>O(6m)$ the r.h.s. of (\ref{c2})
is given by
\begin{eqnarray}
4 \pi r^2 T^r_{\ v}(r>O(6m)) &=& -e^\psi L_H\nonumber\\
-4 \pi r^2 T^v_{\ v}(r>O(6m)) &=& 2 L_H/(1-2m/r)\nonumber\\
4 \pi r T_{rr}(r>O(6m)) &=& 4 L_H/r(1-2m/r)^2\label{cc}\end{eqnarray}

We now estimate $T_{\mu\nu}$ near the horizon by assuming that the
energy momentum tensor measured by an inertial observer falling across the
horizon is finite and of order $L_H$. The jacobian relating $v,r$ to
$(\alpha), (\beta)$ the orthonormal frame of a free falling observer is
$e^\mu_{(0)} = (e^{-\psi}(\Gamma - \dot r)^{-1}, \dot r,
0,0)$,
$e^\mu_{(1)} = (e^{-\psi}(\Gamma - \dot r)^{-1}, \Gamma,
0,0)$ where $
\Gamma = (1- {2 m \over r} + \dot r^2)^{1/2}
$
and $\dot r$ is the rate of change of radius per unit
proper time of the observer. If $\dot r \simeq -1$ near $r=2m$,
the tetrad components are regular because $\Gamma$ does
not vanish. Assuming $T^{(\alpha)(\beta)}=O(L_H/r^2)$
yields
$ r^2 T^r_{\ v}(r\simeq 2m) = O(e^\psi
L_H)$,
$ r^2 T^v_{\ v}(r\simeq 2m) = O(
L_H)$, $ r T_{rr}(r\simeq 2m) = O(L_H / 2 m)$.
The conservation of energy is
\begin{equation}
(r^2 T^r_{\ v})_{,r} + r^2 T^v_{\ v,v} =0
\label{c8}
\end{equation}
where $r^2T^v_{\ v,v} = O(L_{H,v})$. Integrating the conservation equation
from $r=2m$ to $r=O(6m)$ yields $T^r_v$ near the horizon in terms of its
value where (\ref{cc}) is valid. Putting everything together, near the horizon
we have  \begin{eqnarray}  {\partial m
\over \partial v} &=&-L_H e^\psi + O( m L_{H,v})\nonumber\\
{\partial m \over \partial r} &=&O(L_H)\nonumber\\
{\partial \psi\over \partial 2m}
&=&O({L_H / r})\label{c12}
\end{eqnarray}
As anounced all metric coefficients vary slowly if $L_H$
is small and varies slowly. Integrating the equation for $\psi$ yields $e^\psi
\simeq r^{L_H}$ for all $r\geq 2m$. Hence $\psi$ can safely be neglected up to
distances $r = O(e^{1/L_H})$. From now on we suppose for simplicity of the
algebra that $\psi =0$.

Before proceeding we first discuss the location of the horizons in the
geometry (\ref{c1}) with $\psi = 0$. The apparent horizon is the locus where
outgoing geodesics obey $dr/dv =0$, therefore solution of  \begin{equation}
r_a(v) = 2 m (v, r_a(v))\label{c13}
\end{equation}
One may, following Bardeen, define the mass of the black hole at time
$v$ from $M(v) = r_a(v)/2$ whereupon (\ref{c12}) and
(\ref{c13}) give \begin{equation} {d M \over dv} = -L_H + O(L_H^2)+ O(M
L_{H,v})  \label{c14}
\end{equation}
Hence equation (\ref{one}) is recovered provided one proves
$L_H=C/M^2$ and one takes $v$ as time parameter.

We also record here the equation for the event horizon $r_H(v)$. This is the
last light ray which reaches ${\cal I}^+$. It satisfies the equation of
outgoing nul geodesics
\begin{equation}
{dr_H \over dv} = {1 \over 2} { r_H - 2 m(r_H, v)
\over r_H}
\label{c15}
\end{equation}
We obtain an assymptotic expansion for $r_H(v)$ by requiring that it
remains at a finite distance from $r=2M$ for all $v$, ie. neither diverges
to $r=\infty$ nor falls into the singularity at $r=0$.
Setting $r_H (v) = 2 M(v) + \Delta(v)$ we rewrite
(\ref{c15}) in the form $\Delta = 2(2 M + \Delta) ( 2 M_{,v} +
\Delta_{,v}) + 2(m(v, M+\Delta) - M)$. Solving recursively one
obtains a series for $\Delta$ the first term of
which is $\Delta = 8 M M_{,v}$. If one supposes $L_H=C/M^2$ one finds that
$\Delta = -8 C / M$.

In order to calculate the modes or $\langle T_{\mu\nu}\rangle$ we must
first investigate the outgoing radial nul geodesics in the metric (\ref{c1})
with $\psi=0$. It is convenient to change variables to the set $(v, x =
r - r_H(v))$ ie. $x$ is the distance from the
event horizon. In these coordinates the metric becomes
(using (\ref{c15}) )
\begin{equation}
ds^2 = {2 m(v,r_H + x) x\over r_H(r_H + x)} dv^2
-2 dv dx - r^2 d\Omega^2
\label{c16}
\end{equation}
When $x << r_H$  the
equation for radial outgoing nul geodesics can be solved exactly to
yield an exponential approach to the horizon of the form
$\tilde
v - 2 \ln x = f(u)$ where
\begin{equation} \tilde v = \int^v\!dv\
{2m(v,r_H(v))\over r_H^2(v)} \label{c18}\end{equation}
This motivates the following ansatz for the outgoing radial null
geodesics
\begin{eqnarray} \tilde
v - 2 {x\over r_H(v)} -2  \ln x
+ \delta = {u\over 2 \tilde m(u)} + D
\label{c17}\end{eqnarray}
with $D$ a constant of integration.
This should be compared with the solution in the absence of
backreaction
$ v - 2r -4M \ln(r-2M)=u$. The function of $u$ on the r.h.s. of (\ref{c17})
has been writen as $u/\tilde m(u)$ for dimensional reasons. The quantity
$\delta$ is of order  $O(  L_H (Mx+ x^2) / M^2)$ for all $x$ (this is shown by
substitution of (\ref{c17}) into the equation for radial nul geodesics and
integrating the equation for $\delta$ along the geodesics $u=const$). The
function $\tilde m (u)$ is determined by requiring that the variable $u$ in
equation (\ref{c17}) be the same as  in the Vaidya metric equation (\ref{c3}).
The difference $m(u)-\tilde m(u)$ is of order $O(M L_H)$. This is found by
using (\ref{c17}) to change coordinates from the set $(v,r)$ to the set
$(u,r)$ at the radius $r=O(6M)$ where (\ref{c3}) is valid.

Equation (\ref{c17}) is sufficient to prove the first of our hypothesis, to
wit that the flux emitted is $O(M^{-2})$. Indeed, neglecting the potential
barrier and taking only s-waves into account (as in the model of Parentani
and Piran), the flux emitted is given by \cite{DFU}:  $T_{uu}/ 4 \pi r^2 =
(1/12 \pi) (d{\cal U}/du)^{1/2}(d/du)^2(d{\cal U}/du)^{-1/2}$
where ${\cal U}$ labels
the outgoing geodesics as measured by an inertial observer inside the star
(it is analoguous to the Kruskal coordinate $U$). The jacobian $du/d{\cal U}$
is calculated by remarking that at the surface of the star the derivative at
fixed $v$ $d{\cal U}/dr\vert_{v=vstar} = -2$ is constant. Hence  $du/d{\cal
U} = -(1/2) du/ dr\vert_{v=vstar}= (1/2)4 \tilde m(u) / x\vert_{v=vstar}=
-4 \tilde m(u) / {\cal U}$. In this expression we have neglected transients
which occur when the star has just collapsed (these transients vanish as
$e^{-\Delta u/4m}$ where $\Delta u$ is the time since the star began
collapsing). Hence the flux at times $\Delta u > O(4m)$ is equal to $(\pi /
12) T_H^2(u)$ where $T_H(u) = 1/ 8 \pi \tilde m(u) = 1/8 \pi m(u) (1 +
O(L_H))$ as announced. (We mention that their is an intimate relation between
the thermal flux $T_{uu}$ just calculated and the thermal distribution of the
emitted photons. Indeed in the saddle
point evaluation of Bogoljubov coefficients \cite{PB} the
jacobian $du/d{\cal U} = -4 \tilde m(u) / {\cal U}$ is the necessary
ingredient  to prove that the high energy photons
are distributed
Boltzmanly at temperature $T_H(u)$.)

We now return to the second hypothesis and check that $T_{\mu\nu}$ is regular
near the horizon. To this end one writes the Klein Gordon equation in the
coordinates (\ref{c16}) and notes that modes of the form
$\varphi_{\la,l,m}=r^{-1} Y_l^m e^{-i \lambda \tilde
v}  \vert x \vert^{2 i \lambda} \theta (\pm x)$ are
solutions near the horizon $\vert x \vert << M$. The condition of positive
frequency inside the star  is described succintly (see \cite{DR} and
\cite{U}) by asking that one may continue $x$ analytically in the
upper half complex plane. This fixes the ratio of the $\theta(\pm x)$
pieces to be $e^{-2 \pi \lambda}$, hence the thermal character of the flux.
When calculating Green's functions or derivatives thereof at the coincidence
point one  rescales $\lambda$ to $\lambda \to \om = \la 2 m( v , r_H) /
r_H^2$ to obtain the same mode sum in the ultraviolet as in the absence of
back reaction. Therefore the renormalized energy momentum tensor is regular
near the horizon. This is sufficient for our purpose and proves that our
scheme is consistent.

One may however go beyond this qualitative discussion and make a quantitative
estimate of $T_{\mu\nu}$  by using the following adiabatic expansion of the
modes. One writes the solution of the Klein Gordon equation in the absence of
back reaction as $\varphi^0_{\omega,l,m}= r^{-1}Y_l^me^{-i \omega v} \vert
r-2M\vert^{4 i M\omega} f(4M\omega,l, (r-2M)/ 2M)$. Then one
makes the ansatz for the solution in the
metric equation (\ref{c16}) $\varphi_{\la,l,m}=r^{-1} Y_l^m e^{-i \la \tilde v}
\vert x\vert^{ i \la} f(2\la,l, x/ r_H(v)) (1 + \chi(x, v))
$. Calculation shows that $\chi$ is proportional to $L_H$ and encodes
effects such as a change of $O(L_H)$ in the potential barrier in the wave
equation and a little frequency mixing in the potential barrier due to
the time dependence of $m(r,v)$. It gives rise to a modification of the
energy momentum tensor which compared to the estimates in the absence of
back reaction are $\delta T_{\mu\nu} / T_{\mu\nu} = O(L_H)$. The validity of
this adiabatic expansion up to distances $r=O(6 M)$ where the waves propagate
as purely outgoing modes proves that all the emitted quanta, and not only the
Boltzman ones, are distributed thermally with temperature $T_H(u) = 1/ 8 \pi
\tilde m(u)$.

In addition to the Hawking temperature $T_H$ being slightly different from
$1/8 \pi m(u)$ their is also a slight athermicity due to the variation of the
temperature over time. However none of these effects  encode information about
the quantum state of the star. In this respect the semi classical back
reaction resembles the usual Hawking calculation in that one is confronted
with a violation of unitarity for the outside observer\cite{H2}.

We conclude with some remarks concerning the validity of the semi classical
back reaction used so far.
In the preceding calculation we have made appeal
to the geometry and the structure of the quantum field on scales
exponentially  smaller than the Planck scale. This is manifestly
absurd \cite{J}. We now investigate where in space time and at what scale will
quantum back reaction effects completely modify the semi-classical back
reaction. To this end we consider the fluctuations of $T_{\mu\nu}$ and in
particular the energy momentum correlated to the presence (or absence) of a
specific Hawking photon at ${\cal I}^+$. We recall here the main properties
derived in \cite{EMP} of the energy momentum
correlated to the presence of a particular particle (denoted $i_0$) at ${\cal
I}^+$ in a wave packet centered around $u_0$, with energy $\simeq
\omega_0=O(M^{-1})$ : 1) it is given by the conserved tensor $T_{uu}^{i_0}
= (1/4 \pi r^2)f^{i_0}(u)$; 2) $\int\! du\ f^{i_0}(u) = \omega_0$;
3) $f^{i_0}(u)$ takes it values in a region of width $\omega_0^{-1}$
around $u_0$; 4) In addition to the piece of $T_{uu}^{i_0}$ just described in
the outside region $r>r_H$, their is a piece (the partner) on the other side
of the horizon $r<r_H$; 5) Inside the star $T_{\mu\nu}^{i_0}$ takes the
form $(1/4 \pi r^2)f^{i_0}(u)$ after reflecting at $r=0$ whereas before
reflection it is the form  $T_{vv}^{i_0} = (1/4
\pi r^2)f^{i_0}(v)$.

To get a taste for back reaction effects we shall take $T_{\mu\nu}^{i_0}$ as
the source for Einstein's equations. This will give us a qualitative estimate
of when back reaction effects become important. We first consider the
interior of the star. Their one should express distances and energies in
terms of the inertial coordinate ${\cal U}$ rather than $u$, hence reexpress
$T_{uu}^{i_0}$ as $T_{\cal U U}^{i_0} = T_{uu}^{i_0} (du/d{\cal U})^2 =
 T_{uu}^{i_0}(2\tilde m(u_0)/ x\vert_{v=vstar})^2$. The energy density
$(1/4 \pi r^2)T_{\cal U U}^{i_0}$ reaches the Planck scale when
$r=O(1)$ and  $x\vert_{v=vstar} \simeq m(u_0)e^{-u_0/4 \tilde m(u_0)}=
O(1)$and
the quantum field can then no longer be approximated as free in the center of
the star. It is important to notice that the time necessary to reach this
Planckian regime is  $\Delta u = O(m\ln m)$ which should be compared with the
total evaporation time $\Delta u = O(m^3)$.
However this central region can
perhaps be treated in a phenomenological way and could be irrelevant
insomuch as Hawking radiation is considered. Hence we now consider the
fluctuations of the geometry induced by the fluctuations of $T_{\mu\nu}$ near
the horizon. Then integration of equations (\ref{c2}) shows that the effect
of the fluctuation $T_{uu}^{i_0} = (1/4 \pi r^2)f^{i_0}(u)$ induces a change
in the mass parameter $m\to m + O(\omega_0)$ in passing from $u=u_0
-O(\omega_0^{-1})$ to  $u=u_0 +O(\omega_0^{-1})$. However going backwards in
time the photon $i_0$ gets exponentially close to the horizon $r_H$ (since
the light rays (\ref{c17}) adhere exponentially to the horizon). It
is very
soon at a distance $r-r_H = O(M^{-1})$,ie. inside the domain where the horizon
is shifting due to the presence of particle $i_0$ itself. Hence the geometric
concepts of apparent horizon, event horizon, outgoing light rays breaks down
into a quantum soup (the quantum ergosphere of York \cite{Y}) of width  $r-r_H
= O(M^{-1})$. In this region a free field approximation is certainly not
valid. How the quantum theory of gravity contrives to build up (or maybe not
to build up) Hawking radiation out of the soup is the subject of much
conjecture.

\medskip
\noindent
{\bf Acknowledgements.} The author would like to thank R. Brout, F. Englert,
R. Parentani and Ph. Spindel for helpful discussions and R. Brout for a
careful rereading of the manuscript.

 \end{document}